\documentclass{article}%
\usepackage{epsfig}
\usepackage{afterpage}
\usepackage{amsmath}
\usepackage{graphicx}%
\usepackage{amsfonts}%
\usepackage{amssymb}

\DeclareMathAlphabet   {\mathsc}{OT1}{cmr}{m}{sc}

\def\[{\left [}
\def\]{\right ]}
\def\({\left (}
\def\){\right )}
\newcommand{\lang}{\left\langle}
\newcommand{\rang}{\right\rangle}

\newcommand{\oline}[1]{\overline{#1}}

\newcommand{\wh}[1]{\widehat{#1}}

\newcommand{\GeV}      {~\mathrm{GeV}}
\newcommand{\TeV}      {~\mathrm{TeV}}

\newcommand{\EW}       {\mathsc{ew}}

\newcommand{\UV}       {\mathsc{uv}}

\newcommand{\PL}       {\mathsc{pl}}

\newcommand{\GUT}      {\mathsc{gut}}
\newcommand{\STR}      {\mathsc{str}}

\newcommand{\rpart}[1]{{#1}+\oline{#1}}

\newcommand{\hc}       {\mathrm{\; h.c. \;}}

\newcommand{\order}{\mathcal{O}}

\newcommand{\gappeq}{\mathrel{\rlap {\raise.5ex\hbox{$>$}}
{\lower.5ex\hbox{$\sim$}}}}
\newcommand{\lappeq}{\mathrel{\rlap{\raise.5ex\hbox{$<$}}
{\lower.5ex\hbox{$\sim$}}}}

\begin{document}
\begin{center}
            \hfill    MCTP-02-40 \\
            \hfill    hep-ph/0207168
\end{center}

\bigskip
\begin{Large}
\begin{center}
Re-examination of Electroweak Symmetry Breaking in Supersymmetry
and Implications for Light Superpartners
\end{center}
\end{Large}

\bigskip
\begin{center}
G.~L.~Kane{\dag}, J.~Lykken\S, Brent~D.~Nelson\dag\; and
Lian-Tao~Wang\dag
\end{center}

\begin{center}
\dag{\it Michigan Center for Theoretical Physics, Randall Lab.,\\

University of Michigan, Ann Arbor, MI 48109}
\end{center}

\begin{center}
\S{\it Theoretical Physics Department,\\

Fermi National Accelerator Laboratory, Batavia, IL 60510}
\end{center}

\begin{center}
\bigskip ABSTRACT
\end{center}

\begin{quotation}
We examine arguments that could avoid light superpartners as an
implication of supersymmetric radiative electroweak symmetry
breaking. \ We argue that, from the point of view of string theory
and standard approaches to generating the $\mu$ term,
cancellations among parameters are not a generic feature. While
the coefficients relating $M_{Z}$ to parameters in the soft
supersymmetry breaking Lagrangian can be made smaller, these same
mechanisms lead to lighter superpartner masses at the electroweak
scale. \ Consequently we strengthen the implication that gluinos,
neutralinos, and charginos are light and likely to be produced at
the Fermilab Tevatron and a linear collider.
\end{quotation}

\section*{Introduction}
\label{sec:intro}

\qquad One of the main successes of supersymmetry is that it can
provide an explanation of how the electroweak (EW) symmetry is
broken. \ At a general level three assumptions about the form of
the supersymmetric theory are needed for that to work. \ First,
supersymmetry breaking must lead to a soft supersymmetry breaking
Lagrangian with mass terms of order a TeV. \ Second, the $\mu$
term in the superpotential can not be a fundamental product of
Planck-scale physics but must instead be tied to symmetry breaking
at a much lower scale, and third, there must be a quark Yukawa
coupling of order unity. \ The latter two occur naturally in
string theories, so when supersymmetry is viewed as a part of the
four dimensional effective theory following from a more
fundamental string theory they are well motivated. \ The first
must remain an assumption until supersymmetry breaking is
understood.

\qquad For the Higgs potential to actually have a minimum that
breaks the EW symmetry two conditions must be satisfied. \ The one
relevant to us here is the only equation that quantitatively
relates some soft breaking masses at the electroweak scale to a
measured number (at tree level):
\begin{equation}
\frac{M_Z^2}{2} = -\mu^{2}(\EW) + \frac{m^{2}_{H_D}(\EW) -
m^{2}_{H_U}(\EW)\tan^{2}\beta}{\tan^{2}\beta-1} \label{muterm}
\end{equation}
where $m_{H_D}$ and $m_{H_U}$ are the soft masses for the Higgs
doublets coupling to down-type and up-type quarks, respectively,
and $\mu$ is the effective $\mu$ parameter that arises after
supersymmetry breaking (we do not give it a separate name). This
tree level relation can, in turn, be written in the following
way~\cite{KaKi99}
\begin{equation}
M_{Z}^{2}= \sum_{i} C_i m_i^2(\UV) + \sum_{ij}C_{ij}m_i(\UV)
m_j(\UV) \label{zsum}
\end{equation}
Here $m_i$ represents a generic parameter of the softly broken
supersymmetric Lagrangian at an initial high scale $\Lambda_{\UV}$
with mass dimension one, such as gaugino masses, scalar masses,
trilinear A-terms and the $\mu$ parameter.

The coefficients $C_i$ and $C_{ij}$ depend on the scale
$\Lambda_{\UV}$ and quantities such as the top mass and
$\tan\beta$ in a calculable way through solving the
renormalization group equations (RGEs) for the soft supersymmetry
breaking terms. For example, taking the running mass for the top
quark at the Z-mass scale to be $m_{\rm top}(M_Z) = 170 \GeV$, the
starting scale to be the grand-unified scale $\Lambda_{\UV} =
\Lambda_{\GUT} = 1.9 \times 10^{16} \GeV$, and $\tan\beta=5$ we
have for the leading terms in~(\ref{zsum})
\begin{eqnarray}
M_{Z}^{2}&=&-1.8 \mu^{2}(\UV) + 5.9 M_{3}^{2}(\UV) - 0.4
M_{2}^{2}(\UV) -1.2 m^{2}_{H_U}(\UV) \nonumber
\\ & & +0.9 m^{2}_{Q_3}(\UV) + 0.7 m^{2}_{U_3}(\UV)
-0.6 A_{t}(\UV) M_{3}(\UV) \nonumber \\ & & -0.1
A_{t}(\UV)M_{2}(\UV) +0.2 A_{t}^{2}(\UV) +0.4 M_{2}(\UV)M_{3}(\UV)
+ \dots  \label{ztune}
\end{eqnarray}
where the ellipses in~(\ref{ztune}) indicate terms that are less
important quantitatively and for our purposes. In particular
$M_{3}$ and $M_2$ are the $SU(3)$ and $SU(2)$ soft gaugino masses,
respectively, and $A_t$ is the soft trilinear scalar coupling
involving the top squark. \ $C_3$ and $C_{\mu}$, being the largest
coefficients, are those which we will discuss in some detail
below. We think equation~(\ref{zsum}), in a given concrete
manifestation such as~(\ref{ztune}), provides significant insight
into high-scale physics whose implications have not yet been fully
explored.

\qquad Because this equation is the only one connecting
supersymmetry breaking to measured data it was long ago realized
that it was very important~\cite{ElEnNaZw86}-\cite{ChElOlPo99}. \
There is also a connection of supersymmetry to data through the
apparent gauge coupling unification. \ That depends on essentially
the same physics as equation~(\ref{zsum}), requiring the first two
of the three assumptions, but is more qualitative and less able to
tell us precise values for the soft parameters. \ It would be
important if~(\ref{zsum}) could tell us quantitative information
about $M_{3}$ and $\mu$. \ If $M_{3}$ or $\mu$ is large, and the
coefficients $C_3$ and/or $C_{\mu}$ are of order unity or larger,
the right hand side would involve a difference of large numbers,
and explanations in physics don't typically involve such fine
tuning unless a symmetry is present. \ Thus we would tend to
expect both $M_{3}$ and $\mu$ to be rather small. \

\qquad In that case some superpartner masses would be light.
$M_{3}(\UV)$ is closely related to the gluino mass $M_{\tilde{g}}$
-- in the MSSM with $\Lambda_{\UV}=\Lambda_{\GUT}$ we have
$M_{\tilde{g}}\approx 2.9M_{3}(\UV)$ at leading order, with an
increase from squark and gluino loops of as much as $20\%$ -- so
naively one would expect the gluino mass to be small enough so
that it could be observed at the Tevatron. \ Similarly, $\mu$ is
in the chargino and neutralino mass matrices, and if it is small
some charginos and neutralinos should be light enough to be
observed at the Tevatron, even in the limit in which $M_1$ and
$M_2$ are large. \ To be somewhat more precise, for this paper we
will define ``light superpartner'' as one which can be produced at
the Tevatron given its energy and expected luminosity. The actual
definition would depend on the spectrum, but roughly $600 \GeV$ or
less for gluinos and $200\GeV$ or less for the lighter chargino
and neutralinos.

\qquad Under what circumstances might such conclusions fail to
hold? \ There are four arguments that have been suggested. \
First, nature might be unkind and our world may lie at a
particular point in the theory where an accidental cancellation
occurs. \ We cannot prove that is not so, of course. \ But such
accidents are rare in physics, and it is appropriate to proceed on
the assumption that this does not happen. \ Second, there could be
a relation between $M_{3}$ and $\mu$ in the underlying
theory~\cite{ChElOlPo99}, or it could involve some of the other
parameters $m_i$ in the full equation~(\ref{zsum}) with smaller
$C_i$, such as the Higgs mass $m_{H_u}^{2}$ or the other gaugino
masses $M_1$ and $M_2$. \ We will argue below that while not
impossible, this outcome is unlikely. \ Third, the coefficients
from the MSSM could change dramatically in extended theories and,
in particular, become significantly smaller than unity. \ Again,
we will see below that this is unlikely. \ What is more, although
the coefficients can change, physics that reduces the coefficients
(such as beginning the renormalization group evolution of the soft
terms from a lower scale) also tends to reduce the physical
superpartner masses, so the resulting spectrum is not greatly
affected. \ To our knowledge none of these issues have been
systematically addressed previously.

Finally, some ``benchmark models'' for future colliders have been
recently published~\cite{bench,snowmass} which have the systematic
feature of relatively heavy superpartners. \ Because these
benchmark models involve satisfying~(\ref{zsum}) by taking
differences of large numbers, the implication is that either the
MSSM is irretrievably fine-tuned or we must alternatively give up
the explanation of EW symmetry breaking as we understand it in the
context of the MSSM. \ In any event, such models suggest that
there will not be light superpartners observable at the Tevatron.
We will not pursue this in detail in this paper, but in a separate
publication we will exhibit some theoretically well-motivated
models that more naturally satisfy the constraints implied by
equations~(\ref{zsum}) and~(\ref{ztune}) and which have light
superpartners that are observable at the
Tevatron~\cite{benchmark}.

\qquad Historically people first examined equation~(\ref{zsum})
with the assumption of gaugino mass degeneracy, in which case the
large coefficient of $M_{3}$ was taken to also apply for $M_{1}$
and $M_{2},$ leading to predictions of very light charginos and
neutralinos that should have been discovered at
LEP~\cite{BaGi88,deCaMu93b,BaBeOh94}. \ The absence of such
discoveries led to less confidence in the implications of
fine-tuning arguments generally. \ More recently it has been
realized that the coefficients of $M_{1}$ and $M_{2}$, when
treated independently, are quite small and there was no
implication that their discovery was expected at
LEP~\cite{KaKi99}. \ Once $M_1$ and $M_2$ are no longer forced to
be degenerate with $M_3$, electroweak symmetry breaking constrains
them very little. Since $\mu$ is still constrained, however, we
will see that light charginos and neutralinos are still expected.
Indeed, since the coefficient of $\mu$ in equation~(\ref{zsum}) is
always negative and of order one, and since LEP already constrains
$\mu$ to be larger than about $100 \GeV$, electroweak symmetry
breaking inevitably involves some degree of fortuitous
cancellation. Similarly, the large coefficient of $M_{3}$ is still
of some concern even if light gluinos exist; we will address this
issue below.

For completeness, we remark that from the perspective of string
theory it is very natural to have gauge coupling unification but
not gaugino mass degeneracy. Tree level gaugino masses are still
universal, but are often suppressed so that nonuniversal loop
contributions are comparable, while loop contributions to the
gauge couplings themselves are always small in comparison. It is
important to keep this in mind because gaugino mass unification in
conjunction with LEP limits on superpartner masses would imply
large fine-tuning.

\qquad Once superpartners are discovered one might think that the
main implications we are exploring here become less important. \
Further thought shows that this is not so, since
equation~(\ref{zsum}), for a particular choice of $\tan\beta$ and
$\Lambda_{\UV}$ as in~(\ref{ztune}), then becomes a sum rule
giving a constraint on soft breaking parameters which can then
help lead us to a deeper understanding of supersymmetry breaking.

\section{Why $\mu$ and $M_{3}$ are unlikely to be related in the right way}
\label{sec:theory}

\qquad Naively $\mu$ and $M_{3}$ are unlikely to be related
because they seem to arise from different physical mechanisms. \
Supersymmetry breaking generates $M_{3}$, but additional physics
such as the Giudice-Masiero (GM) mechanism~\cite{GiMa88} or a
scalar VEV in the superpotential is needed to generate $\mu.$ \ In
the former case the $\mu$ term vanishes in the absence of
supersymmetry breaking while in the latter it is often associated
with the breakdown of some additional symmetries in the theory
unrelated to supersymmetry in a direct way. Naively, one could
wonder if (say) a stringy approach could produce both a soft
supersymmetry breaking parameter $M_3$ as well as an effective
$\mu$ term in such a way that there is a robust relation between
them that could lead to a cancellation in~(\ref{zsum}) over a
range of parameter space.

\qquad Since there is no compelling theory of the origin of $\mu$
it is not clear how to study this issue. Nevertheless, by
examining these well-established approaches to the $\mu$ problem
in the context of string theory we can understand better what
physics might affect $\mu$ and how it might relate to the physics
that generates gaugino masses. In fact we learn that rather
generally $\mu$ depends on quite different aspects of the theory
from those on which $M_3$ depends. The remainder of this section
is to argue in a number of broad categories that $\mu$ and $M_3$
are largely unrelated.  We will include some detail in the
following so that the non-expert reader can trace the physics.
Readers who accept the previous statement can move to
Section~\ref{sec:scalars}.

\subsection{A first approach: weakly coupled heterotic string
theory}

\qquad Let us first examine the possibility of relating $\mu$ and
$M_3$ in the case of the weakly coupled heterotic string. We
follow the approach of Brignole et al.~\cite{BrIbMu94} in which
one assumes that the communication of supersymmetry breaking from
a hidden sector to the observable sector occurs through the agency
of one of the moduli fields present in string constructions by the
presence of a non-vanishing vacuum expectation value of one or
more of their auxiliary fields $F$. The nature of the soft
supersymmetry breaking terms is then determined by the moduli
couplings to observable sector chiral fields. Here we will employ
invariance under modular transformation as a guide to constructing
these couplings, as in~\cite{BiGaNe01}.

To be completely general, one can allow for both a superpotential
and K\"ahler potential bilinear in the observable sector fields:
\begin{eqnarray}
W(Z^i) &=& \frac{1}{2}\sum_{ij}\nu_{ij}(Z^n)Z^{i}Z^{j}+\dots
\nonumber \\ K(Z^{i}, \oline{Z}^{i}) &=& \sum_{i}
\kappa_{i}(Z^{n})|Z^{i}|^2 +
\frac{1}{2}\sum_{ij}\[\alpha_{ij}(Z^n,
\oline{Z}^{\bar{n}})Z^{i}Z^{j} + \hc\]+\dots. \label{bilinear}
\end{eqnarray}
Here a chiral superfield $Z$ with a superscript $m$, $n$ etc. is
supposed to represent a hidden sector field, and we specifically
have in mind moduli fields. The chiral fields with superscript
$i$, $j$ etc. are observable sector fields. The effective $\mu$
term $\mu_{ij}$ which arises in the superpotential as a result
of~(\ref{bilinear}) for canonically normalized chiral fields
$\wh{Z}^{i} = \kappa_{i}^{-1/2} Z^{i}$, is defined by
\begin{equation}
W(\wh{Z}^{i}) \ni \frac{1}{2}\sum_{ij}\mu_{ij}\wh{Z}^{i}
\wh{Z}^{j} +\hc \label{mudef}
\end{equation}
and this $\mu_{ij}$, which appears in the superpartner mass
matrices, is then given by
\begin{equation}
\mu_{ij} = e^{K/2} (\kappa_i \kappa_j)^{-1/2}\tilde{\mu}_{ij};
\qquad \tilde{\mu}_{ij} = \nu_{ij}
-e^{-K/2}\(\frac{M}{3}\alpha_{ij}-\oline{F}^{\bar{n}}
\partial_{\bar{n}}\alpha_{ij}\).
\label{mueff}
\end{equation}
The terms in~(\ref{mueff}) involving $\alpha_{ij}$ are the result
of the GM mechanism and depend on the auxiliary fields of the
chiral multiplets $\oline{Z}^{\bar{n}}$ and the auxiliary field of
supergravity which is related to the gravitino mass
\begin{equation}
m_{3/2} = -\frac{1}{3} < \oline{M} > = <e^{K/2} \oline{W}>.
\label{gravmass} \end{equation}
We assume throughout, in the manner
of~\cite{BrIbMu94}, that supersymmetry is broken in such a way as
to ensure zero vacuum energy at the minimum of the scalar
potential.

Modular invariance of the expressions in~(\ref{bilinear}) implies
particular functional forms for the $\nu_{ij}$ and $\alpha_{ij}$
which depend on the modular weights of the fields $Z^i$ involved.
Assuming for maximum simplicity that these functions have no
dependence on the dilaton $S$ and that they are both holomorphic
in the (overall) modulus field $T$, then modular invariance of the
K\"ahler potential and covariance of the superpotential
in~(\ref{bilinear}) requires~\cite{BiGaNe01}
\begin{equation}
\alpha_{ij}(Z^n) = [\eta(T)]^{-2(n_i + n_j)}; \qquad \nu_{ij}(Z^n)
=\left[ \eta(T) \right]^{-2(3+n_i+n_j)}, \label{mod}
\end{equation}
where $\eta(T)$ is the Dedekind eta function and $n_i$ is the
modular weight of the field $Z^i$, defined by $\kappa_i = (T +
\oline{T})^{n_i}$.

Using the simple tree level K\"ahler potential $K=-\ln(\rpart{S})
- 3\ln(\rpart{T})$ for the moduli we find in the pure
Giudice-Masiero case ($\nu_{ij} = 0$)
\begin{equation} \mu =
m_{3/2}(\rpart{t})^{-(n_{H_u} + n_{H_d})/2}[\eta(t)]^{-2(n_{H_u} +
n_{H_d})} \label{GMmu}
\end{equation}
whereas the properly normalized, tree level gaugino mass, under
the same assumptions, is given by
\begin{equation}
M_{a}=\frac{g_{a}^{2}(\Lambda_{\rm UV})}{2} F^{S} +
\Delta_{a}^{\rm loop} = \frac{g_{a}^{2}(\Lambda_{\rm
UV})}{g_{\STR}^{2}(\Lambda_{\STR})}\sqrt{3}m_{3/2}\sin\theta +
\Delta_{a}^{\rm loop}. \label{gluinoSUGRA}
\end{equation}
The second equality holds when the vacuum energy vanishes and we
have taken $<s+\bar{s}> = 2/g_{\STR}^2$ where $g_{\STR}$ is the
unified gauge coupling at the string scale
$\Lambda_{\STR}$.\footnote{It is quite common in the literature to
assume that $\Lambda_{\UV} = \Lambda_{\GUT} = \Lambda_{\STR}$, and
to ignore possible threshold effects in the determination of the
running couplings $g_{a}^{2}(\Lambda)$, in which case we recover
the familiar results of ``minimal supergravity'' as found
in~\cite{BrIbMu94} with a single universal gaugino mass.} Here
$\theta$ is the familiar Goldstino angle of~\cite{BrIbMu94},
parameterizing the degree to which the two moduli $S$ and $T$
participate in supersymmetry breaking, with $\theta=0$ implying
moduli domination $(F^{S} = 0)$ and $\theta=\pi/2$ implying
dilaton domination $(F^{T} = 0$). We have represented the
nonuniversal contributions that arise at one loop by
$\Delta_{a}^{\rm loop}$. In string models $\Delta_{a}^{\rm loop}$
will generally be a function of auxiliary fields for the various
moduli in the theory as well as the auxiliary field of
supergravity~\cite{GaNeWu99,BaMoPo00}. These contributions are
crucial in creating nonuniversalities in the gaugino sector, but
for the sake of our argument in this section we will assume
$\Delta_{a}^{\rm loop} = 0$ and merely ask whether $\mu$ can be
related to the {\em tree} level value of $M_3$. \ Although both
$\mu$ and $M_{3}$ \ vanish when $M_{3/2}$ does, they depend
generically on very different physics and it is not reasonable to
assume they are related over a region of the parameters.

\qquad The example given above represents the simplest possible
outcome from a string perspective as this is sufficient to see the
lack of correlation between $M_3$ and $\mu.$\footnote{More
complicated cases can be found in the literature, such as allowing
the function $\alpha_{ij}$ to arise through a nonrenormalizable
operator that depends on fields charged under an anomalous
$U(1)$~\cite{DuGrPoSa96,MoRi97}.} Alternatively we can examine the
more involved and well-known calculation of $\mu$ by Antoniadis et
al.~\cite{AnGaNaTa94} whose phenomenology was studied
in~\cite{BrIbMu96,BrIbMuSc97}. Bilinear terms in the K\"ahler
potential were identified for (2,2) compactifications in which a
richer moduli spectrum is present. To be as specific as possible,
we assume such terms are present for the Higgs fields of the MSSM,
which have kinetic terms of the form
\begin{equation}
\kappa_{H_u} = \kappa_{H_d} = \frac{1}{(T + \oline{T})(U +
\oline{U})} \label{kappaHiggs}
\end{equation}
where now $T$ is meant to represent one of the three K\"ahler
moduli and $U$ is a single complex structure modulus. Under the
modified modular symmetries that arise in models with continuous
Wilson lines~\cite{AnGaNaTa94,CaLuMo94} the invariant bilinear
term is given by
\begin{equation}
\alpha_{H_u H_d} =  \frac{1}{(T + \oline{T})(U + \oline{U})} .
\label{alphaHiggs}
\end{equation}
Then the $\mu$ term is obtained from~(\ref{mueff})
\begin{equation}
\mu= \[m_{3/2}-\frac{\oline{F}^T}{(T + \oline{T})}
-\frac{\oline{F}^U}{(U + \oline{U})}\] \to
m_{3/2}\[1-\sqrt{3}\cos\theta(\Theta_T+\Theta_U)\] \label{muANT}
\end{equation}
where the Goldstino angle $\theta$ now distinguishes the dilaton
from the combined $T$ and $U$ sector and $\Theta_T$ and $\Theta_U$
represent the individual contributions to SUSY breaking from the
particular moduli fields participating in the bilinear
term~(\ref{alphaHiggs})~\cite{BrIbMu96,BrIbMuSc97}. Given that the
gluino mass continues to take the form of~(\ref{gluinoSUGRA}), it
is very unlikely that the various moduli VEVs and the pattern of
supersymmetry breaking represented by the angles $\theta$,
$\Theta_T$ and $\Theta_U$ would take precisely the necessary
values to maintain a relationship between $\mu$ and $M_{3}$ that
could account for the smallness of $M_{Z}.$

\qquad The above examples both employed a bilinear in the K\"ahler
potential in the spirit of Giudice and Masiero. \ One could
instead assume that the $\mu$ term arises from a more complicated
effective quadratic term in the superpotential represented by
$\nu_{ij}$ in~(\ref{bilinear}). For example, consider a higher
order nonrenormalizable term in the superpotential where an
effective $\mu$ term is to be generated through the vacuum value
of a sequence of fields. Schematically we imagine a superpotential
of the form
\begin{equation}
W(Z^{i}) \sim \frac{Z^{n_1}Z^{n_2}\dots Z^{n_k}}{M_{\PL}^{k-1}}
H_u H_d \to \Lambda_{X} \(\frac{\Lambda_X}{M_{\PL}}\)^{k-1} H_u
H_d, \label{Wstring}
\end{equation}
where the latter expression is the effective $\mu$ term when the
fields take a VEV $<Z> \sim \Lambda_X$. If this scale can be
engineered to be an intermediate scale such as $10^{11} \GeV$ then
a dimension four term would suffice to generate an effective $\mu$
of about the electroweak scale. This was the basic idea behind the
work of~\cite{KiNi84,ChKiNi92}.\footnote{Though in that case the
bilinear quantity receiving a VEV was assumed to be a
quark-antiquark bilinear $Q\oline{Q}$ from the hidden sector, so
it represents a nonrenormalizable operator stemming from the
low-energy field theory as opposed to one originating from the
underlying string theory, where we would expect a holomorphic term
such as the one in~(\ref{Wstring}).} Alternatively, if we imagine
the fields $Z^i$ in~(\ref{Wstring}) to originate from higher order
terms in the string compactification then we might image the scale
$\Lambda_{X}$ to be the scale of anomalous $U(1)$ breaking. Then
these VEVs would be much larger since typically
$(\Lambda_{X}/m_{\PL}) \sim 1/10$ and an electroweak-scale $\mu$
term would require an operator of rather large dimension.

For anomalous $U(1)$ breaking we expect the combination of fields
cancelling the Fayet-Illiopoulos term to preserve modular
invariance~\cite{GaGi02}. Therefore the canonically normalized
effective $\mu$ term is given by
\begin{equation}
\mu = v(\frac{v}{m_{\PL}})^{k-1} \frac{[\eta(t)]^{-2(3+n_{H_u} +
n_{H_d})}}{\((\rpart{s})(\rpart{t})^{(3+n_{H_u} +
n_{H_d})}\)^{1/2} } \label{anomMU}
\end{equation}
with $v$ representing a moduli-dependent generic vacuum value of
the scale of the anomalous $U(1)$ breaking. Again, $\mu$ depends
on physical quantities quite different from those $M_{3}$ depends
on, since $M_3$ would continue to be given by~(\ref{gluinoSUGRA}).

We might alternatively imagine that the $\mu$ term arises from a
renormalizable (dimension three) term in the superpotential such
as $W_N = \lambda_{H_u H_d N}(T) N H_u H_d$, with $N$ a singlet
under the gauge groups of the Standard Model. Such a term is often
embedded in the so-called ``Next-to-Minimal'' Supersymmetric
Standard Model (NMSSM) where it is assumed to be accompanied by a
term such as $W_k = kN^3$~\cite{ElGuHaRoZw89,KiWh95}. However,
terms such as $W_k$ are hard to come by in string
constructions~\cite{CvLa96}. Since we are thinking here of
possible trilinear terms that are likely to be of fundamental
origin from the string theory point of view, we might instead
imagine $W_N$ coming from an effective $E_6$ inspired model as
might be expected in Calabi-Yau
compactifications~\cite{ElEnNaZw86,Ri89,CvDeEsEvLa97,ClCvEsEvLa98},
in which case $W_N$ is accompanied by additional trilinear terms
coupling the Standard Model singlet to exotic states such as
vector like triplets of SU(3) $D$ and $\oline{D}$. The details of
how a vacuum value $<\lambda_{H_u H_d N}(T) N>\; \simeq 1 \TeV$ is
arranged is immaterial for our purposes, however. It is sufficient
to note that if the Yukawa coupling $\lambda_{H_u H_d N}$ is
allowed by the string selection rules (which will often be the
case since such a coupling is allowed by $E_6$ gauge invariance
where $N$ is interpreted as a singlet under the SO(10) subgroup of
$E_6$), then it is presumably a fundamental (tree level) string
term and modular invariance of the superpotential term $W_N$ is to
be expected. We are thus led to an effective $\mu$ term
\begin{equation}
\mu = \lambda_{H_u H_d N}<N> \frac{[\eta(T)]^{-2(3+n_{H_u} +
n_{H_d}+ n_N)}}{\((\rpart{s})(\rpart{t})^{(3+n_N + n_{H_u} +
n_{H_d})}\)^{1/2}}, \label{E6mu} \end{equation} where the constant
$\lambda_{{H_u}{H_d}N}$ is presumably $\order(1)$. Again, the
gluino mass is still given by~(\ref{gluinoSUGRA}). In both of
these superpotential examples the magnitude of the effective $\mu$
is related to the breaking of some symmetry (an anomalous $U(1)$
in the first example, a $U(1)'$ arising from the breaking of $E_6$
to the Standard Model in the second) not directly related to
supersymmetry. Even if the initial challenge of arranging these
symmetry-breaking scales properly to ensure $\mu \sim
\order(M_{3/2})$ can be overcome, we still are no better off than
in the K\"ahler potential cases studied previously.

\subsection{More general string theory examples}

So far the discussion has been couched in the context of weakly
coupled heterotic string theory compactified at a high scale
$\Lambda_{\STR} \simeq M_{\PL}$, with supersymmetry breaking
transmitted via moduli to the observable sector at a slightly
lower scale $\Lambda_{\UV} \simeq \order(10^{16}-10^{17}) \GeV$.
Nevertheless our conclusions continue to hold in the
strongly-coupled heterotic limit of M-theory in the absence of
nonperturbative objects such as five-branes. In such cases the
tree level gaugino masses are modified from their relation
in~(\ref{gluinoSUGRA}) to the
form~\cite{NiOlPo97,ChKiMu98,LuOvWa98b,CeMu99}
\begin{equation}
M_a =\frac{g_{a}^{2}(\Lambda_{\rm
UV})}{g_{\STR}^{2}(\Lambda_{\STR})}
\frac{\sqrt{3}m_{3/2}}{1+\epsilon}\[\sin\theta +
\frac{1}{\sqrt{3}} \epsilon \cos\theta\] \label{HWgaugino}
\end{equation}
where $\epsilon$ is a parameter which is zero in the weak-coupling
limit and we have suppressed the possible relative phase between
the two terms. We continue to expect the $\mu$ term to be
generated by one of the mechanisms stated above, as was considered
in~\cite{CeMu99}.

Meanwhile recent developments in string theory have led to a great
deal of research on Type I/Type IIB string theories, with
fundamental string scales that can be significantly lower than the
Planck scale. This activity has thus far provided few new
insights, however, into the origin of the $\mu$ term and its
relation to soft supersymmetry breaking parameters such as $M_3$.
For example, phenomenological studies of D-brane models
constructed directly from string theory, with an intermediate
string scale $\Lambda_{\STR} \sim 10^{11} \GeV$, have generally
left $\mu$
unspecified~\cite{BuIbQu99,IbMuRi99,AbAlIbKlQu00,CeGaKhMuTo01},
allowing us to pick the mechanism above that suits our fancy. As
these examples continue to imagine SUSY breaking communicated from
a hidden brane to our observable brane via moduli, we can continue
to use the framework of~\cite{BrIbMu94} to explore the soft terms
in such theories.

For example, in the class of models studied by Iba\~nez, Mu\~noz
and Rigolin~\cite{IbMuRi99} we find that the precise form of the
gluino mass will now depend on the type of brane with which the
$SU(3)$ gauge group of the Standard Model is associated. If it is
associated with a D9-brane the gluino mass continues to be given
by the familiar~(\ref{gluinoSUGRA}), where $\theta$ continues to
distinguish the dilaton sector from three distinct untwisted
moduli fields $T^{i}$.\footnote{In the various embeddings
considered in~\cite{CeGaKhMuTo01}, for example, this was always
the form of the gluino mass.} Should $SU(3)$ be associated with
one of three possible sets of D5-branes, however, the gluino mass
would be \begin{equation} M_{3}=\frac{g_{3}^{2}(\Lambda_{\rm
UV})}{2} F^{T_i} = \frac{g_{3}^{2}(\Lambda_{\rm
UV})}{2}\sqrt{3}m_{3/2}\Theta_i\cos\theta , \label{gluinoDbrane}
\end{equation}
where the parameter $\Theta_i$ determines the degree to which the
moduli of the $i$th D5-brane contributes to the cancellation of
the vacuum energy, with $\sum_{i} \Theta_{i}^{2} = 1$. In fact, if
we allow for the likely presence of various twisted moduli in the
theory then we expect expressions such as~(\ref{gluinoSUGRA})
and~(\ref{gluinoDbrane}) to involve yet more Goldstino
angles~\cite{BuIbQu99,IbMuRi99}, thus further depressing any hope
of obtaining a robust relationship between $\mu$ and $M_3$.

\qquad Thus we see very generally that a relation between $\mu$
and $M_{3}$ that gives a non-accidental cancellation seems very
unlikely in string-derived models with intermediate to high string
scales. \ This conclusion is strengthened by keeping in mind that
the required cancellation is really a function of such things as
Yukawa couplings and $\tan\beta$, as these parameters govern the
renormalization group evolution of these parameters, and hence
determine the coefficients such as those in~(\ref{ztune}).
Furthermore, in this section we have restricted ourselves to the
simplest and most conservative formulae, ignoring effects such as
expected gaugino mass non-degeneracies at the loop level or the
presence of additional moduli which participate in supersymmetry
breaking, which would only strengthen our conclusions.

Ideally we would survey all models in the literature in an attempt
to confirm that $M_3$ and $\mu$ are never related in just such a
way as to have small fine tuning in the electroweak sector. Even
apart from the difficulty of such a task, many models of very
low-energy strings or brane-world constructions are not formulated
with sufficient precision to judge their fine-tuning implications.
Often it is considered sufficient to merely obtain $\mu \sim
m_{3/2} \sim m_{1/2}$ to avoid fine-tuning and hence be considered
``natural.'' However it is difficult to imagine such an imprecise
relation guaranteeing small cancellations in~(\ref{zsum})
and~(\ref{ztune}) over a range of the free parameters in such
models, as we have explained in the context of more precise cases
examined above. To put it differently, if there exists a model in
which certain soft parameters and $\mu$ are related {\em of
necessity} in just the needed manner as to have negligible amounts
of fine-tuning in the electroweak sector, then this model should
be taken very seriously. We now turn to a broader consideration of
the coefficients of $M_{3}$ and $\mu$ to understand their size and
theoretical model-dependence.

\section{Why $M_3$ is unlikely to be related to other soft terms in the right
way} \label{sec:scalars}

It may appear more reasonable to require a cancellation between
the gluino mass $M_3(\UV)$ and the up-type Higgs mass
$m_{H_U}^{2}(\UV)$ at the input scale than a relation between
$M_3$ and the $\mu$ parameter. The main conclusion to be drawn
from the previous section is that, at a minimum, the value of
$M_Z$ depends on two unrelated scales: the gross scale of the soft
supersymmetry breaking terms, characterized by $m_{3/2}$, and the
scale of symmetry breaking that determines $\mu$. Even when these
two gross scales are related, as in the mechanism of Giudice and
Masiero, the details of the physics determining $M_3$ and $\mu$
suggest that obtaining relative magnitudes for these parameters
that are related in such a way as to allow for their absolute
magnitudes to be large without concomitant large cancellations
would be an accidental outcome of an underlying theory.

But assuming, for the moment, that such a theory is found -- can
we still find relations {\em among} the soft terms in the
Lagrangian themselves such that a heavy gaugino sector need not
imply large cancellations in~(\ref{zsum})? After all, the soft
Lagrangian is grossly defined by only one parameter, $m_{3/2}$.
For example, the simplest and most constrained limit of the weakly
coupled heterotic string is the so-called ``dilaton dominated''
limit in which only the dilaton plays a role in transmitting
supersymmetry breaking from the hidden sector to the observable
sector. In that case we most certainly do have a relation between
gaugino and scalar masses, namely $M_a = \sqrt{3} m_{0}$, where
both the scalar masses and gaugino masses are unified at the GUT
scale. And this is a ``robust'' relation in the sense that it does
not depend on certain other parameters in the theory, such as the
vacuum values of string moduli like $S$ and $T$.

So relations among soft terms are more likely to occur than
between $\mu$ and $M_3$, but are they likely to be both robust and
also of the sort to allow large values for these soft terms,
relative to $M_Z$, without large cancellations? To investigate
this equation we need to look beyond the gross features of the
soft terms and consider the details of their dependence on the
underlying theory, as we did in Section~\ref{sec:theory}. But
first let us consider the special role played by $m_{H_U}^{2}$ in
determining the Z mass.

\subsection{Relating soft gaugino masses to $m_{H_U}^{2}$}

We first consider the special role in equation~(\ref{zsum}) played
by $m_{H_U}^{2}$ in determining the Z mass. The coefficient
$C_{H_U}$ is both of the right sign and general magnitude to
provide cancellation against $C_3$. But note that in the case of
universal scalar masses this feature disappears. Taking all scalar
masses in~(\ref{ztune}) to be given by $m_0$ we would have
\begin{equation}
M_Z^2 = -1.8 \mu^2 (\UV) + 0.4 m_{0}^{2} + 5.9 M_{3}^{2} - 0.4
M_{2}^{2} + \dots \label{mSUGRA}
\end{equation}
so this mechanism of making large values of $M_3$ natural requires
nonuniversal scalar masses. In particular the Higgs masses must be
divorced from the scalar masses of the matter fields of the MSSM.
This is not surprising as $m_{H_U}^2$ has a privileged place in
the soft Lagrangian: its running to negative values triggers the
EW symmetry breaking that gives rise to nonzero values of $M_Z$ in
the first place.\footnote{In our analysis of Section~\ref{sec:RGE}
below we will also see that the magnitude of $C_{H_U}$ is also
unique in that its coefficient remains quite constant, even in
extended theories. This is a manifestation of the same physics
behind the ``focus point'' effect in its running, as noted
in~\cite{FeMaMo00a}.}

Nor is it impossible to imagine, though to do so we must
investigate the structure of these soft terms in greater detail.
The tree level gaugino masses and scalar masses in a general
supergravity theory are given by
\begin{eqnarray} M^{0}_a &=&  \frac{g_a^2}{2} F^n
\partial_n f_{a}^{0} . \label{gaugtree} \\
(m^{0}_{i})^2&=& \lang \frac{M\oline{M}}{9} - F^n
\oline{F}^{\bar{m}}\partial_n\partial_{\bar{m}}\ln\kappa_i \rang ,
\label{scalartree}
\end{eqnarray}
where $f_a^0 = S$ in the weakly-coupled limit, $M$ is the
supergravity auxiliary field from~(\ref{gravmass}) and $\kappa_i$
was defined in~(\ref{bilinear}). Tree level nonuniversality could,
in principle, arise from differing moduli-dependence of the
K\"ahler metrics $\kappa_i$ of the Higgs fields from the rest of
the observable sector, say via differing modular weights $n_i$.
Alternatively, we could imagine cases in which the tree level
scalar mass in~(\ref{scalartree}) is precisely zero, as in models
with a no-scale structure. Then the leading scalar masses arise
through loop effects and will necessarily be nonuniversal. Each of
these outcomes is possible, but the theory must now arrange for
the contributions to each of the individual scalar masses, as well
as the gluino mass, to be of such a magnitude as to allow {\em
automatic} cancellations among these soft terms -- and thereby
allow these parameters to be much larger than $M_Z$ without
fine-tuning through large cancellations. The question again
becomes model-dependent.\footnote{For an analysis of fine-tuning
in the case of nonuniversal Higgs soft masses,
see~\cite{ChChNa98}.}

From the string theory perspective, even in the simplest case of
the weakly-coupled heterotic string, the soft Lagrangian is in
general determined by {\em three} independent scales as opposed to
simply $m_{3/2}$. The first of these is the overall scale of
supersymmetry breaking given by $M \sim m_{3/2}$ and appearing as
the leading term in the soft scalar masses~(\ref{scalartree}). The
other two scales depend on the degree to which the two broad
classes of string moduli (the dilaton and the compactification
moduli such as the $T$ and $U$) participate in the transmission of
this supersymmetry breaking to the observable sector, via nonzero
auxiliary fields $F$. The gaugino masses will be determined, at
leading order in weak coupling, by the dilaton contribution. By
contrast, scalar masses feel the effects of moduli auxiliary
fields through the moduli dependence of the $\kappa_i$, which tend
to include the compactification moduli. This suggests that each of
these three scales must be engineered to be of the right
proportions to one another. Then we return to the case of $\mu$
and $M_3$ where the smallness of $M_Z$ can be maintained in the
face of large values for the soft supersymmetry breaking gaugino
masses, without the implication of large cancellations
in~(\ref{zsum}), only provided certain precise model-dependent
relations hold.

\subsection{Relating gaugino masses to one another}
\label{sec:gaugino}

Perhaps we would find better success by restricting our attention
to the gaugino sector alone. In this case we would be working with
just one overall scale to leading order -- that defined by
$<F^S>$. Thus we might expect relations between gaugino masses to
more easily arise as robust (parameter-independent) predictions
from string theory than relations between gaugino masses and other
soft terms. To illustrate with an example, in the case
of~(\ref{mSUGRA}), should a model predict $M_2 \simeq 17 M_3$ then
the common scale (given, say, by the vacuum value of the dilaton
auxiliary field) becomes essentially irrelevant: gaugino masses of
any magnitude will not imply large cancellations in the
determination of the Z-mass. Note that if only partial
cancellation occurs then we may indeed allow for larger gaugino
mass terms for a given degree of fine-tuning, but we are still
lead to the general conclusion that gluinos must be light if
electroweak symmetry breaking is not accidental. The
model-independent objects that might give rise to a robust
relation between $M_2$ and $M_3$ include such things as the ratio
of gauge couplings $g_{2}^{2}(\mu)/g_{3}^{2}(\mu)$ or
beta-function coefficients $b_2/b_3$, where
\begin{equation}
b_a = \frac{1}{16\pi^2} \( 3 C_a - \sum_i C_a^i \),  \label{ba}
\end{equation}
and $C_a$, $C_a^i$ are the quadratic Casimir operators for the
gauge group ${\cal G}_a$, respectively. While neither of these
objects are likely to produce factors of $\order(10-20)$ in the
context of the MSSM, it is conceivable that theories which provide
robust relations between the gauginos of the right magnitude can
be constructed.

To summarize, cancellations of some sort must clearly occur
in~(\ref{zsum}) in order to obtain $M_Z = 91 \GeV$. This is not in
dispute. Furthermore, every reasonable model of physics at the
supersymmetry breaking scale must make predictions for the soft
terms and $\mu$ which will imply certain relations among them that
depend on internal model dynamics. But the larger each individual
term becomes in a particular expression such as~(\ref{ztune})
or~(\ref{mSUGRA}), the more precisely those internal parameters
must be specified to avoid an unreasonable cancellation in the
determination of the Z mass. Alternatively, the smaller the
individual soft terms are, in particular the value of $\mu$ and
$M_3$, the less we must rely on such precise relations and the
less ``accidental'' $M_Z = 91 \GeV$ becomes.

\section{The coefficients of $\mu$ and $M_{3}$}
\label{sec:RGE}

Accepting the premise, then, that the observed magnitude of the Z
mass is not an accidental outcome of an underlying theory, we now
seek to investigate the strength of this argument to changes in
the framework used to obtain expressions such as~(\ref{ztune}).
Since the coefficients (fine-tuning parameters) in~(\ref{zsum})
are obtained from the input parameters by solving the
renormalization group equations (RGEs) and running from the input
scale down to the electroweak scale, their values could
potentially depend on the assumptions we make in solving the RGEs.
We now study the effects of changing some of those assumptions. In
what follows we will work with one loop RGEs and solve the tree
level electroweak symmetry breaking conditions. Including the full
one loop radiative corrections to the Higgs potential tends to
reduce the fine-tuning, particularly for small $\tan\beta$, but
will not change the conclusions we will draw in this section.

\subsection{The dependence on the choice of $\tan\beta$ and
$\lambda_{\rm top}$} \label{sec:tanbeta}

\begin{figure}
\begin{center}
\includegraphics[scale=0.35,angle=270]{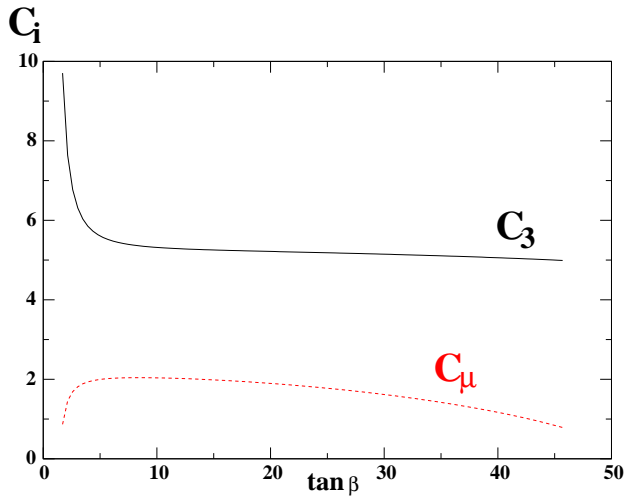}
\epsfxsize=15cm
\end{center}
\caption{\bf The dependence of the fine-tuning coefficients $C_3$
and $C_{\mu}$ on the value of $\tan\beta$ for a running top quark
mass at $M_Z$ of $m_{\rm top}(M_Z) = 164 \GeV$.
\label{finetune_tanbeta}}
\end{figure}

One of the key ingredients of radiative electroweak symmetry
breaking is the existence of a Yukawa coupling of $\order(1)$. For
a large region of the parameter space, the large Yukawa coupling
is provided by the top quark Yukawa $\lambda_t$. Indeed, the
nature of the electroweak symmetry breaking ({\em i.e.} the values
of the fine-tuning coefficients $C_i$) is quite sensitive to the
choice of $\lambda_t$, as is well
known~\cite{BaGi88,deCa93,CaOlPoWa94,DiGi95}. Of course this
choice depends on the value of $\tan\beta$ and $\lambda_t$. We
studied the effect of $\lambda_t$ on the fine-tuning coefficients
$C_i$ for the leading terms $C_3$ and $C_{\mu}$.
Figure~\ref{finetune_tanbeta} plots the values of these two
coefficients as a function of $\tan\beta$ for a running top mass
at the Z-mass of $m_{\rm top}(M_Z)$ of $164 \GeV$ and
$\Lambda_{\UV}$=$\Lambda_{\GUT}$. The values of $C_3$ and
$C_{\mu}$ are most sensitive at low values of $\tan\beta$ and are
largely independent of $\tan\beta$ for $\tan\beta
\stackrel{>}{\sim} 4$ and below its perturbative
limit.\footnote{Another fine-tuning issue arises for large values
of $\tan\beta$, but that is not directly related to our concerns
here.}

\begin{figure}
\begin{center}
\includegraphics[scale=0.35,angle=270]{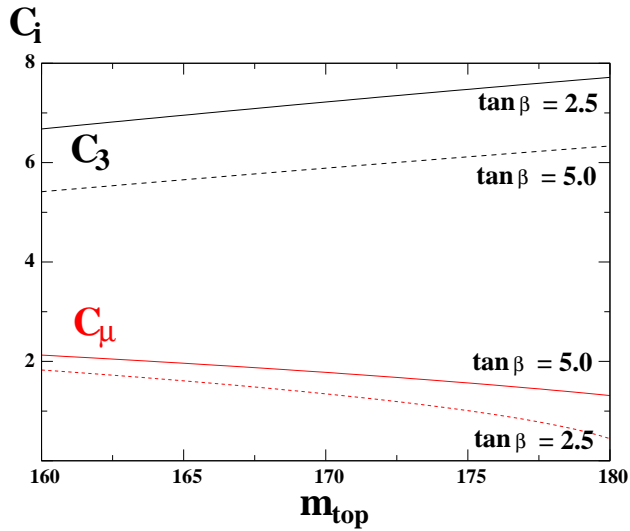}
\epsfxsize=15cm
\end{center}
\caption{\bf The dependence of the fine-tuning coefficients $C_3$
and $C_{\mu}$ on the value of the running top quark mass $m_{\rm
top}(M_Z)$, for $\Lambda_{\UV} = 3 \times 10^{16} \GeV$.
\label{finetune_mt}}
\end{figure}

In Figure~\ref{finetune_mt} we choose two values of $\tan\beta$
and investigate the importance of $m_{\rm top}$ on the same
coefficients for an initial scale $\Lambda_{\UV} =
\Lambda_{\GUT}$. It has long been appreciated that fine-tuning
related to the $\mu$ parameter was relaxed for larger values of
$m_{\rm top}$ and larger values of $\tan\beta$, as is borne out by
Figures~\ref{finetune_tanbeta} and~\ref{finetune_mt}. What is not
often appreciated is that these same directions tend to {\em
increase} the fine-tuning related to the gluino mass $M_3$. This
complementarity is clearly evident in
Table~\ref{tbl:coefficientsTOP}, where the coefficients $C_i$ and
leading coefficients $C_{ij}$ of~(\ref{zsum}) are given, for an
array of values for the running top mass at $M_Z$ and $\tan\beta$
with an initial scale of $\Lambda_{\UV}=\Lambda_{\GUT}$.

\begin{table}[thb]
{ {\begin{center}
\begin{tabular}{|l|c|c|c|c|c|} \cline{1-6}
  & Case 1.A & Case 1.B & Case 1.C & Case 1.D & Case 1.E \\ \cline{1-6}
 $\tan\beta$ & 2.5 & 2.5 & 35 & 5 & 2.5 \\
 $m_{\rm top} (M_Z)$ & $165\GeV$ & $174\GeV$ & $170\GeV$ & $170\GeV$ &
 $170\GeV$ \\
 \cline{1-6}
 $M_{1}^{2} (\UV)$ & 0.02 & 0.03 & -0.004 & 0.001 & 0.02\\
 $M_{2}^{2}(\UV)$ & -0.2 & -0.2 & -0.4 & -0.4 & -0.2\\
 $M^{2}_{3}(\UV)$ & 6.9 &  7.4 & 5.4 & 5.9 & 7.1\\
 \cline{1-6}
 $A^{2}_{\rm top}(\UV)$ & 0.2 & 0.1 & 0.2 & 0.2 & 0.1\\
 $A^{2}_{\rm bot}(\UV)$ & 0 & -0.0006 & -0.02 & -0.0008 & -0.0006\\
 $A^{2}_{\rm tau}(\UV)$ & 0 & 0 & 0.0004 & 0 & 0 \\
 \cline{1-6}
$\mu^{2}(\UV)$ & -1.7 & -1.2 & -1.5 & -1.8 & -1.4
\\ \cline{1-6} $m^{2}_{Q_3}(\UV)$ & 1.0 & 1.2 & 0.8 & 0.9 & 1.1 \\
$m^{2}_{U_3}(\UV)$ & 0.8 & 0.9 & 0.7 & 0.7 & 0.9 \\
$m^{2}_{D_3}(\UV)$ & 0.07 & 0.07 & 0.03 & 0.06 & 0.07 \\
$m^{2}_{L_3}(\UV)$ & -0.07 & -0.07 & -0.05 & -0.06 & -0.07
\\ $m^{2}_{E_3}(\UV)$ & 0.07 & 0.07 & 0.05 & 0.06 & 0.07 \\
$m^{2}_{H_u}(\UV)$ & -1.3 & -1.2 & -1.2 & -1.2 & -1.3 \\
$m^{2}_{H_d}(\UV)$ & 0.3 & 0.3 & -0.08 & 0.03 & 0.3 \\
 \cline{1-6}
$M_2(\UV)M_3(\UV)$ & 0.5 & 0.4 & 0.4 & 0.4 & 0.4 \\ $A_t(\UV)
M_3(\UV)$ & -0.6 & -0.4 & -0.6 & -0.6 & -0.5
\\ $A_t(\UV) M_2 (\UV)$ & -0.1 & -0.1 & -0.1 & -0.1 & -0.1 \\ \cline{1-6}
\end{tabular}
\end{center}}
{\footnotesize \caption{{\bf Coefficients $C_i$ and leading
$C_{ij}$ for $M_Z$ as a function of running top mass at $M_Z$ and
$\tan\beta$}. } \label{tbl:coefficientsTOP}} }
\end{table}

\subsection{The dependence on the choice of input scale}
\label{sec:lambdaUV}

The usual assumption is to run the RGEs from the scale where the
Standard Model gauge couplings unify, $\Lambda_{\GUT}$. However,
it is typical that supersymmetry breaking actually appears in the
observable sector at a lower scale, such as $\Lambda_{\rm INT}
\sim 10^{11} - 10^{14} \GeV$ in models with intermediate string
scales, or $\Lambda_{\rm GMSB} \sim 10^{5}-10^{8} \GeV$ in
gauge-mediated models. In these cases the input scale for the soft
parameters $\Lambda_{\UV}$ should be identified as the lower
supersymmetry breaking scale. More generally, without a concrete
model of supersymmetry breaking (and indeed, often even in cases
with one) it is somewhat uncertain where to initiate RG evolution
of the soft parameters and what sorts of corrections to include.
We will see that lowering the input scale can have sizable effects
on the values of some of the fine-tuning coefficients.

\begin{table}[thb]
{ {\begin{center}
\begin{tabular}{|l|c|c|c|c|c|} \cline{1-6}
  & Case 2.A & Case 2.B & Case 2.C & Case 2.D & Case 2.E \\
\cline{1-6}
 $\Lambda_{\UV} ({\rm GeV})$ & $1 \times 10^{16}$ & $1 \times 10^{14}$ &
 $1 \times 10^{11}$ & $1 \times 10^{8}$ & $1 \times 10^{5}$ \\
 \cline{1-6}
 $M_{1}^{2} (\UV)$ & 0.001& -0.006 & -0.01 & -0.02 & -0.01 \\
 $M_{2}^{2}(\UV)$ &  -0.4 & -0.3 & -0.2 & -0.3 & -0.2 \\
 $M^{2}_{3}(\UV)$ &  5.9 & 4.4 & 2.6 & 1.3 & 0.4
 \\ \cline{1-6}
 $A^{2}_{\rm top}(\UV)$ & 0.2 & 0.2 & 0.2 & 0.3 & 0.2 \\
 $A^{2}_{\rm bot}(\UV)$ & -0.0008 & -0.0006 & -0.0006 & -0.0004 & -0.0002\\
 $A^{2}_{\rm tau}(\UV)$ & 0 & -0.00006 & -0.00004 & -0.00004 & 0 \\
 \cline{1-6}
$\mu^{2}(\UV)$ & -1.8 & -1.8 & -1.7 & -1.7 & -1.8
\\ \cline{1-6} $m^{2}_{Q_3}(\UV)$ & 0.9 & 0.8 & 0.7 & 0.6 &
0.4
\\ $m^{2}_{U_3}(\UV)$ & 0.7 & 0.7 & 0.6 & 0.5 & 0.4
\\ $m^{2}_{D_3}(\UV)$ & 0.06 & 0.05 & 0.04 & 0.02 & 0.01 \\
$m^{2}_{L_3}(\UV)$ & -0.06 & -0.05 & -0.04 & -0.02 & -0.01
\\ $m^{2}_{E_3}(\UV)$ & 0.06 & 0.05 & 0.04 & 0.02 &
0.01
\\ $m^{2}_{H_u}(\UV)$ & -1.2 & -1.3 & -1.4 & -1.5 & -1.7
\\ $m^{2}_{H_d}(\UV)$ & 0.03 & 0.03 & 0.05 & 0.06 & 0.07
\\ \cline{1-6}
$M_2(\UV)M_3(\UV)$ & 0.4 & 0.3 & 0.2 & 0.08 & 0.02 \\ $A_t(\UV)
M_3(\UV)$ & -0.6 & -0.6 & -0.5 & -0.4 & -0.2
\\ $A_t(\UV) M_2 (\UV)$ & -0.1 & -0.1 & -0.1 & -0.07 & -0.03 \\ \cline{1-6}
\end{tabular}
\end{center}}
{\footnotesize \caption{{\bf Coefficients $C_i$ and leading
$C_{ij}$ for $M_Z^2$ in~(\ref{zsum}) as a function of
$\Lambda_{\UV}$ for $M_{\rm top} = 170 \GeV$ and $\tan\beta=5$}. }
\label{tbl:coefficientsUV}} }
\end{table}

In Table~\ref{tbl:coefficientsUV} we display the coefficients
$C_i$ and leading $C_{ij}$ of~(\ref{zsum}) as a function of the
input scale $\Lambda_{\UV}$, where we have set $m_{\rm top}(M_Z) =
170 \GeV$ and $\tan\beta = 5$ throughout. Note that the
coefficient $C_{\mu}$ in front of $\mu^2(\UV)$ is essentially
independent of the input scale while the coefficient $C_{H_u}$
grows with lower input scale -- eventually supplanting $C_3$ as
the most important soft mass in determining $M_Z$. Given that the
sign of $C_{H_u}$ and $C_{\mu}$ are the same,
Table~\ref{tbl:coefficientsUV} suggests that provided
$m_{H_u}^{2}(\UV)$ is positive at the initial SUSY breaking scale
the overall problem of unnatural cancellations in~(\ref{zsum}) is
worse in lower-scale SUSY breaking mechanisms.

\subsection{The dependence on possible matter at intermediate
scales} \label{sec:intscales}

Usually, studies of radiative electroweak symmetry breaking are
based on the assumption that the MSSM is valid from the
electroweak scale $\Lambda_{\EW}$ all the way up to $\Lambda_{\UV}
\simeq \Lambda_{\GUT}$. On the other hand, many models constructed
from some fundamental theory contain a number of fields in
addition to the MSSM matter content. While it is quite possible
that all of these fields decouple at $\Lambda_{\UV}$, in which
case they would have no effect on the discussion of fine-tuning
described above, it is also plausible that some of them may
acquire masses at some intermediate scale below the SUSY breaking
scale $\Lambda_{\EW}<\Lambda_I<\Lambda_{\UV}$. The intermediate
scale matter will then give threshold corrections which alter the
running of the soft parameters and can affect the values of the
corresponding fine-tuning coefficients.

Let us consider this question in a general context before
investigating specific cases. Clearly the importance of the gluino
mass $M_3$ in~(\ref{zsum}) is a manifestation of its importance in
the RG evolution of the soft parameters entering
into~(\ref{muterm}). How might the presence of additional charged
matter affect this evolution? Consider for a moment the system of
equations given in schematic form by
\begin{eqnarray}
\frac{d m^2_{H_U}}{dt}& \sim & \frac{1}{16 \pi^2}
\[|\lambda_t|^2(m_{H_u}^2 + m_Q^2 + m_U^2)  + |A_{t}^{2}|^2\] \cdots,
\label{rgemhu}
\\ \frac{d \lambda_t}{dt}  & \sim &  \frac{\lambda_t}{16 \pi^2}
\[|\lambda_t|^2 - \frac{16}{3} g_3^2 \]+ \cdots , \label{rgeyu} \\
\frac{d A_t}{d t} &\sim& \frac{1}{16 \pi^2}
\[A_t(|\lambda_t|^2 - \frac{16}{3} g_3^2) + \lambda_t
(\frac{32}{3} g_3^2 M_3 )\] +\cdots, \label{rgeau} \\ \frac{d
m_{\tilde{q}}^2}{dt}
 & \sim & \frac{1}{16 \pi^2} \[ m_{\tilde{q}}^2
|\lambda_t|^2  -\frac{32}{3} g_3^2 |M_3|^2\]  +\cdots,
\label{rgemq}
\end{eqnarray}
where we have have kept only the leading terms proportional to
top-quark Yukawas or $g_3$. We have also introduced the running
scale $t=\ln (\Lambda/M_Z)$. In equation~(\ref{rgemq}) $\tilde{q}$
could be any scalar quark soft mass.

A few comments are in order. The dependence of $m_{H_U}^2(\EW)$ on
$M_3^2(\UV)$ comes from the strong $M_3^2$ dependence of the
running of soft parameters such as $m_{\tilde{q}}^2$ and $A_t$.
For the scalar masses this dependence always appears in the form
\begin{equation} g_3^2 |M_3|^3 = g_3^6(t)
\frac{M^2_3(\UV)}{g^4_3(\UV)}, \label{invariant}
\end{equation}
where we have factored out $\frac{M^2_3(\UV)}{g^4_3(\UV)}$ since
it is an RGE invariant at the leading order at which we are
working. Adding intermediate scale matter then will affect the RGE
evolution of the above quantities, and hence the fine-tuning
coefficient of $M_3$, provided the intermediate scale particles
are charged under color. Not only will such particles give
threshold corrections to the running of $g_3$, directly feeding
into the $M_3^2$ term in the RGEs, but the running of the Yukawa
coupling $\lambda_t$ is also largely controlled by the $g_3^2$
term in its RGE. Therefore, changing the running of $g_3^2$ will
have a large effect on the value of $\lambda_t^2(\UV)$ which in
turn has a significant impact on the fine-tuning coefficients.

The intermediate scale matter can also be charged under $SU(2)$
and couple to MSSM up-type Higgs with a new Yukawa coupling. If
such a coupling exists it will lead to extra terms in the RGE of
$m_{H_U}^2$. Let $X_q$ represent new chiral fields which are
doublets under $SU(2)$ and triplets under $SU(3)$ and let $X_l$
represent new chiral superfields which are doublets under $SU(2)$
alone. Then if couplings with $H_u$ exist we expect new terms in
the RGE for $m_{H_u}^2$ of the heuristic form
\begin{equation}
\delta\(\frac{d m^2_{H_U}}{dt}\) = \theta_{Xq}(t)|\lambda_{Xq}|^2
m^2_{X_q} + \theta_{X_l}(t) |\lambda_{X_l}|^2 m_{X_l}^2.
\label{intrgemhu}
\end{equation}
The $\theta$'s in~(\ref{intrgemhu}) are properly defined step
functions turning on the couplings above the energy scale
$\Lambda_I$. We must distinguish, then, the intermediate matter
based on their charge under $SU(3)$ versus $SU(2)$. The running of
any new ``squarks'' $m^2_{X_q}$ will necessarily introduce new
positive contributions into the running of $m_{H_u}^{2}$ due to
the $M_3^2$ term in its RGE. This will in turn enhance the
fine-tuning parameter $C_3$.\footnote{Notice that a sizable Yukawa
coupling between some heavy fields charged under color to the
ordinary Higgs is the necessary and sufficient condition for an
$\order(1)$ enhancement of the fine-tuning. It does not depend on
the initial value of $m^2_{X_q}$ because the running of
$m^2_{X_q}$ will contain a large contribution of $M_3(\UV)$
regardless of its initial value.} On the other hand, adding new
``leptons'' $m_{X_l}^2$ as in the second term in~(\ref{intrgemhu})
will not bring in large new contributions to the running of
$m_{H_u}^{2}$ because their RGEs are independent of the gluino at
leading order. An immediate conclusion one can draw from the
discussion here is that if there is additional matter at some
intermediate scale charged under color with a Yukawa coupling to
the MSSM Higgs, those Yukawa couplings will be strongly
constrained by the naturalness requirement of the electroweak
symmetry breaking.

Next, we turn to study the effects of intermediate scale matter on
the value of $C_{\mu}$. At one loop, the RGE of $\mu$ in the MSSM
is
\begin{equation}
\frac{d \mu}{dt} = \frac{\mu}{16 \pi^2} \[3|\lambda_t|^2 + 3
|\lambda_b|^2 + |\lambda_{\tau}|^2 - 3g_2^2 - \frac{3}{5} g_1^2
\].
\label{muRGEint}
\end{equation}
First, we notice that since $\mu$ only receives wavefunction
renormalizations, the solution of its RGE will always have the
form $\mu(t)= \mu(\UV) f(t,\dots)$, where $f$ is a function of
scale and other parameters, and therefore $C_{\mu} \sim |f|^2$.
Positive terms on the right hand side of~(\ref{muRGEint})
proportional to the square of the magnitudes of Yukawa couplings
will therefore cause $C_{\mu}$ to grow and negative ones
proportional to the squares of the gauge couplings will cause it
to diminish. This is the origin of the observed behavior of the
coefficients $C_3$ and $C_{\mu}$ in the context of the MSSM:
stronger running of $g_3$ is correlated with lower $C_3$ but
larger $C_{\mu}$.

The existence of potential intermediate scale matter can have
several types of effects on the running of $\mu$, above and beyond
the threshold corrections to the beta functions of the gauge
couplings, however. If the intermediate scale matter couples to
the MSSM Higgs through some Yukawa type couplings $\lambda_X$ then
there will be an additional effect on the running of $\mu$. Those
corrections always have the form of \begin{equation} \delta
\(\frac{d\mu}{dt}\) = \theta(t) \sum_X |\lambda_X|^2,
\end{equation}
where $\theta(t)$ is zero below the intermediate scale $\Lambda_I$
and $\sum_X \lambda_X$ represents the sum of all such new Yukawa
couplings.\footnote{We will later be interested in particular
Yukawa couplings which do not couple to fields charged under color
since otherwise they will enhance the fine-tuning significantly.
An example might include a Yukawa coupling between a heavy
right-handed neutrino to a left-handed neutrino and Higgs
providing the off-diagonal elements in the see-saw mass matrix. }
Generically, we see that the inclusion of extra states with
non-vanishing Yukawa couplings to the Higgs sector can make
$C_{\mu}$ smaller.

The study of the solutions of the RGEs above helps to illuminate
the origin of the fine-tuning coefficients. Next we will use this
insight to study the possibility of suppressing the fine-tuning
parameters $C_3$ and $C_{\mu}$ through judicious choices of
intermediate scale matter combinations and couplings. We begin
with the change of the beta functions of the gauge couplings from
the introduction of intermediate scale matter. The beta function
coefficients of the MSSM, in their GUT normalization, are given by
\begin{equation}
\frac{dg_a}{dt} = - \frac{b_a g_{a}^{3}}{16\pi^2}; \qquad
b_1=-\frac{33}{5}; \;\; b_2=-1; \;\; b_3=3.
\end{equation}
Above the intermediate scale $\Lambda_I$ they will be modified to
$b'_i=b_i+ \delta_i$. For simplicity let us initially choose
$\delta_1=\delta_2=\delta_3=\delta$, which would represent new
matter in complete multiplets of $SU(5)$ which preserves gauge
coupling unification at the usual GUT scale $\Lambda_{\GUT} \simeq
2 \times 10^{16} \GeV$.

We can achieve a reduction in $C_3$ (at the expense of a higher
value of $C_{\mu}$ as mentioned above) by putting in a negative
$\delta$ through the addition of new fundamentals charged under
the Standard Model. To be concrete we can take as an example the
states coming from a chiral supermultiplet of representations
$\mathbf{27}$ and $\mathbf{\overline{27}}$ of $E_6$. Under the
decomposition under the Standard Model gauge group,\footnote{The
numbers in the parenthesis label the representations under $SU(2)$
and $SU(3)$ while the lower indices outside are the hypercharges.}
those states are~\cite{MaRa95}
\begin{enumerate}
\item $({\mathbf{2,1}}^c)_{-1} + c.c.$ \qquad $\delta_i=(-\frac{3}{5},-1,0)$
\item $({\mathbf{1,1}}^c)_{2} + c.c.$ \qquad $\delta_i=(-\frac{6}{5},0,0)$
\item $({\mathbf{1,\bar{3}}}^c)_{-\frac{4}{3}} + c.c.$ \qquad
$\delta_i=(-\frac{8}{5},0,-1)$
\item $({\mathbf{1,\bar{3}}}^c)_{\frac{2}{3}} + c.c.$ \qquad
$\delta_i=(-\frac{2}{5},0,-1)$
\item $({\mathbf{2,3}}^c)_{\frac{1}{3}} + c.c.$ \qquad
$\delta_i=(-\frac{1}{5},-3,-2)$
\end{enumerate}

If, for example, we take one copy of them to be at some
intermediate scale, the total threshold correction will be
$\delta_1=\delta_2=\delta_3= \delta= -4$. We can also have states
coming from vector supermultiplets of the adjoint representation
of $E_6$. They decompose similarly under the Standard Model gauge
group as the first five of the states coming from the chiral
supermultiplet. The threshold correction of one copy of those
states will be $\delta=12$.

As we mentioned above, the inclusion of this type of threshold
correction will induce some enhancement in the fine-tuning
coefficient $C_{\mu}$ associated with $\mu$.  In
Table~\ref{tbl:coefficientsINT} we demonstrate this fact by
presenting the complete set of coefficients $C_i$ and the leading
$C_{ij}$ for different sets of intermediate scale matter,
including a case where the additional fields do not come in
complete multiplets of $SU(5)$ but are instead designed to yield
gauge coupling unification at the string
scale~\cite{MaRa95,GaNe00a}. We have begun the running of all soft
terms, as well as the $\mu$ parameter, from the scale
$\Lambda_{\UV}$ which we have taken to coincide with the scale of
gauge coupling unification in each case. The values in
Table~\ref{tbl:coefficientsINT} do not assume any additional
Yukawa couplings between the new matter multiplets and the fields
of the Standard Model. As a consequence the reduction in $C_3$ is
always related to an increase in $C_{\mu}$.

\begin{table}[thb]
{
{\begin{center}
\begin{tabular}{|l|c|c|c|} \cline{1-4}
 & Case 3.A & Case 3.B & Case 3.C \\ \cline{1-4}
 $\Lambda_{\UV} ({\rm GeV})$ & $3 \times 10^{16}$ & $3 \times 10^{16}$ & $4 \times 10^{17}$ \\
 $\Lambda_{\rm int} ({\rm GeV})$ & $5 \times 10^{11}$ & $5 \times 10^{11}$ & $1 \times 10^{14}$\\
 $\alpha_{\GUT} (\Lambda_{\UV})$ & 0.29 & 0.06 & 0.05\\
 \cline{1-4}
 $M_{1}^{2} (0)$ & 0.003 & -0.0002 & 0.0008\\
 $M_{2}^{2}(0)$ & -0.05 & -0.2 & -0.3 \\
 $M^{2}_{3}(0)$ & 0.6 & 3.3 & 4.3 \\
 $A^{2}_{\rm top}(0)$ & 0.2 & 0.2 & 0.2 \\
 $A^{2}_{\rm bot}(0)$ & -0.0007 & -0.0007 & -0.0008 \\
 $A^{2}_{\rm tau}(0)$ & 0 & 0 & 0 \\
 \cline{1-4}
$\mu^{2}(0)$ & -2.9 & -2.2 & -2.2 \\ $m^{2}_{Q_3}(0)$ & 0.8 & 0.8
& 0.8
\\ $m^{2}_{U_3}(0)$ & 0.6 & 0.6 & 0.6
\\ $m^{2}_{D_3}(0)$ & 0.07 & 0.06 & 0.06 \\
$m^{2}_{L_3}(0)$ & -0.07 & -0.06 & -0.06 \\ $m^{2}_{E_3}(0)$ &
0.07 & 0.06 & 0.06
\\ $m^{2}_{H_u}(0)$ &-1.3 & -1.3 & -1.4 \\
$m^{2}_{H_d}(0)$ & 0.008 & 0.02 & 0.02 \\ \cline{1-4}
$M_2(\UV)M_3(\UV)$ & 0.07 & 0.3 & 0.4 \\ $A_t(\UV) M_3(\UV)$ &
-0.3 & -0.6 & -0.6
\\ $A_t(\UV) M_2 (\UV)$ & -0.1 & -0.1 & -0.2 \\ \cline{1-4}
\end{tabular}
\end{center}}
{\footnotesize \caption{{\bf Coefficients $C_i$ for $M_Z^2$ as
in~(\ref{zsum}) for various intermediate matter scenarios}. Case
(A) has a common delta of -12. Case (B) has a common delta of -4.
Case (C) has $\delta_1 =-1$, $\delta_2 = -3$ and $\delta_3 = -4$,
which represents the case where we have added two pairs of
$(D,\oline{D})$ and one pair of $(Q,\oline{Q})$. This table
assumes a top mass $m_{\rm top}(M_Z) = 164 \GeV$ and $\tan\beta =
5$. } \label{tbl:coefficientsINT}} }
\end{table}

\begin{figure}
\begin{center}
\includegraphics[scale=0.35,angle=270]{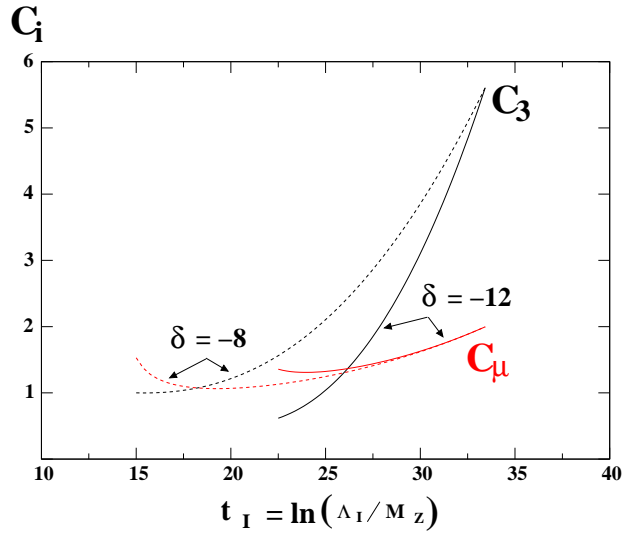}
\epsfxsize=15cm
\end{center}
\caption{{\bf The dependence of the fine-tuning coefficients $C_3$
and $C_{\mu}$ on threshold corrections and the location of the
intermediate scale $\Lambda_I$.} The solid curves are for a common
$\delta = -12$ while the dashed curves are for a common
$\delta=-8$. This plot is for $\tan\beta = 5$, $m_{\rm top}(M_Z) =
164 \GeV$ and assumes a new set of Yukawas such that $\sum_X
|\lambda_X|^2 = 5$. \label{finetune_tI}}
\end{figure}

However, this increase in $C_{\mu}$ can be mitigated by
introducing some extra Yukawa couplings associated with some heavy
field, as in~(\ref{muRGEint}), while being sure not to allow
sizable Yukawa interactions between the new heavy fields which are
charged under color as this will induce new contributions to the
fine-tuning of $M_3$ through the RGE of $m^2_{H_U}$. In
Figure~\ref{finetune_tI} we show the resulting reduction in both
$C_3$ and $C_{\mu}$ with two different values of $\delta$. For
both cases we assume a set of additional Yukawa couplings
$\lambda_X$ between the new heavy color-singlet states that appear
at the scale $\Lambda_I$ and the Higgs fields of the Standard
Model. For concreteness we have chosen $\sum_X |\lambda_X|^2 =5$
so as to maximize the effect.

A simultaneous reduction in both $C_3$ and $C_{\mu}$ which
preserves gauge coupling unification would require a special
combination of exotic states with (a) some being charged under
$SU(3)$ of the Standard Model to reduce $C_3$ but no Yukawa
couplings to the MSSM Higgs sector and (b) additional states which
are singlets under $SU(3)$ with generally large Yukawa couplings
to the Higgs sector to reduce $C_{\mu}$.

\subsection{But can the gluino be heavy after all?}
\label{sec:heavygluino}

One might take the analysis in Sections~\ref{sec:tanbeta} -
\ref{sec:intscales}, which shows that the relation between SUSY
breaking parameters and $M_Z$ is model dependent, as evidence that
the fine-tuning of equation~(\ref{zsum}) can be reduced or removed
in some theories. But in the simple examples we have discussed
here that is not so -- put simply, whatever physics reduces the
size of $C_3$ or $C_{\mu}$ through an alteration of the RG
evolution of $M_3$ and $\mu$ will simultaneously change the
predicted superpartner masses such that they remain light. For
example, the case described in Figure~\ref{finetune_tI} shows a
possible reduction in $C_3$ by a factor of almost 10, depending on
the scale $\Lambda_I$ at which the intermediate matter is
introduced. However, the physical gluino mass in the modified
theory is smaller by an amount such that any naturalness bound
implies about the same gluino mass in all cases. Explicitly, for
the MSSM unified at $2 \times 10^{16} \GeV$ one has $M_3(\EW)
\approx 3 M_3(\UV)$, while for the maximum reduction in
Figure~\ref{finetune_tI} one has $M_3(\EW) \approx 0.43 M_3(\UV)$.

\begin{table}[thb]
{
{\begin{center}
\begin{tabular}{|l|c|c|c|c|c|} \cline{1-6}
 Case & $\Lambda_{\UV} (\GeV)$ & $\Lambda_{\rm int} (\GeV)$ & $C_3$ &
 $M_3 (\UV)$ & $M_3(\EW)$ \\ \cline{1-6}
Case 2.A & $1 \times 10^{16}$ & NA & 5.9 & $84 \GeV$ & $241 \GeV$
\\ Case 2.B & $1 \times 10^{14}$ & NA & 4.4 & $98 \GeV$ & $251 \GeV$
\\ Case 2.C & $1 \times 10^{11}$ & NA & 2.6 & $126 \GeV$ & $274 \GeV$
\\ Case 2.D & $1 \times 10^{8}$ & NA & 1.3 & $181 \GeV$ & $323 \GeV$
\\ Case 2.E & $1 \times 10^{5}$ & NA & 0.4 & $340 \GeV$ & $474 \GeV$
\\ \cline{1-6}
Case 3.A & $3 \times 10^{16}$ & $5 \times 10^{11}$ & 0.6 & $263
\GeV$ & $106 \GeV$
\\ Case 3.B & $3 \times 10^{16}$ & $5 \times 10^{11}$ & 3.3 & $113 \GeV$
& $232 \GeV$
\\ Case 3.C & $4 \times 10^{17}$ & $1 \times 10^{14}$ & 4.3 & $98 \GeV$
& $236 \GeV$
\\
 \cline{1-6}
\end{tabular}
\end{center}}
{\footnotesize \caption{{\bf Electroweak scale running gluino mass
for equivalent cancellations in~(\ref{zsum})}. The last column
column shows the value of $M_3(\EW)$, which is the observed gluino
pole mass $M_{\tilde{g}}$ up to loop corrections, for various
examples of reduced $C_3$ studied in Sections~\ref{sec:lambdaUV}
and~\ref{sec:intscales}.}  \label{tbl:gluinomass}} }
\end{table}

We can make this phenomenon even more concrete by revisitng the
various examples in Tables~\ref{tbl:coefficientsUV}
and~\ref{tbl:coefficientsINT} and ask what the implications of
varying $C_3$ are for the low-scale gluino mass. For illustrative
purposes let us say that the individual terms in~(\ref{zsum}) are
individually no larger than $5M_{Z}^{2}$; for other choices the
conclusion is not substantially changed.\footnote{This choice is
roughly equivalent to requiring a ``sensitivity'' parameter
$\Delta_3 \equiv |(M_3/M_{Z}^{2})\partial M_{Z}^{2}/\partial M_3|
\leq 10$, which in earlier days would have been considered an
``upper bound'' on tolerable amounts of
fine-tuning~\cite{BaGi88,DiGi95}.} In particular this choice
implies that for each case considered in
Tables~\ref{tbl:coefficientsUV} and~\ref{tbl:coefficientsINT} we
must have
\begin{equation}
M_3 (\UV) = \sqrt{5/C_3}M_Z \label{M3solve}. \end{equation} In
Table~\ref{tbl:gluinomass} we have collected the eight cases and
computed the implied $M_3(\UV)$ value as well as the resulting
electroweak scale value after RG evolution to the scale of $M_Z$
under the assumptions of each scenario. Even in the most extreme
cases considered here a light gluino seems inescapable if we are
to avoid large cancellations in~(\ref{zsum}).

There is, however, a possible loophole. One can imagine a UV
theory in which there is a fixed nonuniversal UV relation between
$M_1$, $M_2$, and $M_3$, such that effectively the coefficient of
$M_3$ is greatly reduced without any other accompanying ``side
effects.'' As we discussed in Section~\ref{sec:gaugino}, such a
theory would allow significantly larger gluino masses for a given
degree of fortuitous cancellations. Obviously this can only happen
if the UV gaugino mass relations make $M_1$ and/or $M_2$ {\it
larger} than $M_3$ at the UV scale. This is the opposite behavior
to what happens in gauge mediation, anomaly mediation, and other
models where nonuniversal gaugino masses are related to the
corresponding gauge couplings or beta function coefficients. Thus
to exploit this loophole one is forced into a rather
unconventional theoretical corner, perhaps utilizing the
suggestions of~\cite{Anderson}.

\section*{Concluding comments}

\qquad Our goal here has been to take seriously and study
phenomenologically the implications of assuming that supersymmetry
breaking leads to, and indeed, when the origin of $\mu$ is
included, explains electroweak symmetry breaking. \ Those
implications are not theorems. \ In a sense our analysis provides
evidence for our tentative conclusions. In this paper we are
essentially taking a top-down approach to understanding
electorweak symmetry breaking -- we study how the parameters
originate in the theory to better understand their roles and
implications.

\qquad The analysis of the relation between $\mu$ and $M_{3}$ that
arises from deriving electroweak symmetry breaking, and thus
$M_Z$, strongly disfavors the possibility of a robust cancellation
between them. A similar cancellation among soft parameters is also
disfavored. This implies that it is indeed appropriate to
interpret equations such as~(\ref{ztune}) as implying that the
gluino and the parameters which appear in the chargino and
neutralino mass matrices should all be truly $\order(M_Z)$. Even
allowing for a factor of a few to account for possible accidental
cancellation it is still the case that in the MSSM gluinos should
be at most a few times $M_{Z}$ and some charginos and neutralinos
lighter than that. With the MSSM values of $C_{3}$ and $C_{\mu}$
the implied upper limits on gaugino masses are quite small, though
not inconsistent with experimental constraints.

\qquad The implications of studying how the coefficients can vary
in extended theories are also subtle. \ First, in any reasonable
approach within the framework of the MSSM alone, one or more of
the coefficients $C_i$ remain larger than unity, implying some
light superpartners exist (always with the caveat that electroweak
symmetry breaking is not a coincidence based on accidental
cancellations). \ Perhaps the concerns of the previous paragraph
could be partly mitigated because the coefficients, particularly
$C_{3}$, were actually reduced by the possible presence of
intermediate scale matter below the supersymmetry breaking scale,
or simply by beginning the RG evolution of supersymmetry-breaking
parameters from a lower scale. \ But as we saw in the simple
approaches considered here, the connection between high and low
scale sparticle masses changes in a way correlated with the
changes in $C_3$, so the actual gluino, chargino, and neutralino
masses are not likely to be larger than in the MSSM case.

Once the soft parameters are measured one can use the implied sum
rule embodied in~(\ref{zsum}) as a tool in reconstructing an
underlying theory of supersymmetry breaking. Computing sets of
coefficients as in Tables~\ref{tbl:coefficientsUV}
and~\ref{tbl:coefficientsINT}, representing potential new physics
between the scale of supersymmetry breaking and the electroweak
scale, may suggest that certain patterns of high-scale soft terms
produce more reasonable cancellations than others. \ In the
meantime, the re-examination of the implications of the
supersymmetric radiative electroweak symmetry breaking strengthens
the likelihood that superpartners are being produced at the
Fermilab Tevatron if supersymmetry indeed provides the explanation
of EW symmetry breaking. \ Similarly, a $500 \GeV$ linear collider
should then be above the threshold to produce lighter charginos
and neutralinos.

We can summarize the implications as follows. Several
phenomenological clues point to light superpartners, but
electroweak symmetry breaking gives our best quantitative
constraints. If weak scale supersymmetry is a correct idea, some
UV theory generated the soft breaking  and $\mu$ parameters at
some higher energy scale. These parameters then resulted in
electroweak symmetry breaking, with the measured mass of the $Z$
as an output. We would prefer to believe that the smallness of
$M_Z$ is not mostly due to large accidental cancellations of
parameters which are essentially unrelated at the UV matching
scale. As we have shown, this either requires rather light
gluinos, charginos, and neutralinos, or it requires a UV theory
with special features unlike what we observe in existing
frameworks, both stringy and non-stringy.

What would we conclude if superpartners are not detected at the
Tevatron? There are four possibilities, the first being that SUSY
is simply not the explanation of electroweak symmetry breaking.
Only slightly more palatable is to conclude that rather large
accidental cancellations actually do occur, or to conclude that
the UV theory enforces rather odd relations between $M_3$ and
other parameters (most probably $M_1$ and/or $M_2$). The most
acceptable conclusion is likely to be that the Tevatron {\it did}
produce superpartners, but that they were not detected, due to
degradation of the discovery signatures as happens in a large
number of fairly conventional SUSY scenarios, or to insufficient
luminosity to reconstruct a signal. Of course this only
re-emphasizes the importance of pursuing aggressive superpartner
searches at the Tevatron. If superpartners are not observed at the
Tevatron after considerable integrated luminosity, it is
interesting to think about what arguments imply they should be
observed at the LHC.

\section*{Acknowledgements}

We appreciate discussions with, and assistance from,
P.~Bin\'etruy, T.~Dent, L.~Everett, J.~Giedt, T.~Wang, J.~Wells
and E.~Witten.

\end{document}